\documentclass[twocolumn,showpacs,preprintnumbers,amsmath,amssymb,groupedaddress,pre]{revtex4}

\usepackage{graphicx}
\usepackage{dcolumn}
\usepackage{bm}
\usepackage{epsfig}

\def\e{\begin{equation}}
\def\f{\end{equation}}
\def\=#1{\overline{\overline #1}}

\def\-#1{{\bf #1}}
\def\.{\cdot}
\def\l#1{\label{eq:#1}}
\def\r#1{(\ref{eq:#1})}

\begin{document}

\title{Sub-wavelength imaging at infrared frequencies using an array of metallic nanorods}

\author{M\'ario G. Silveirinha}
\affiliation{Departamento de Engenharia Electrot\'{e}cnica da
Universidade de Coimbra, Instituto de Telecomunica\c{c}\~{o}es,
P\'{o}lo II, 3030 Coimbra, Portugal}

\author{Pavel A. Belov}
\affiliation{Department of Electronic Engineering, Queen Mary
University of London, Mile End Road, London, E1 4NS, United Kingdom}
\affiliation{Photonics and Optoinformatics Department, St.
Petersburg State University of Information Technologies, Mechanics
and Optics, Sablinskaya 14, 197101, St. Petersburg, Russia}

\author{Constantin R. Simovski}
\affiliation{Photonics and Optoinformatics Department, St.
Petersburg State University of Information Technologies, Mechanics
and Optics, Sablinskaya 14, 197101, St. Petersburg, Russia}

\date{\today}

\begin{abstract}
We demonstrate that an array of metallic nanorods enables
sub-wavelength (near-field) imaging at infrared frequencies. Using
an homogenization approach, it is theoretically proved that under
certain conditions the incoming radiation can be transmitted by the
array of nanorods over a significant distance with fairly low
attenuation. The propagation mechanism does not involve a resonance
of material parameters and thus the resolution is not strongly
affected by material losses and has wide bandwidth. The
sub-wavelength imaging with $\lambda/10$ resolution by silver rods
at 30 THz is demonstrated numerically using full-wave
electromagnetic simulator.
\end{abstract}

 \maketitle

\section{Introduction}

Recently, the problem of sub-diffraction imaging has attracted a lot
of interest in the scientific community. In 2000, J. Pendry has
shown that a left-handed material slab with negative index of
refraction makes a perfect lens with unlimited resolution
\cite{Pendry_PerfectLens}. The key property that enables
sub-diffraction imaging in Pendry's lens is the amplification of the
evanescent spatial harmonics. This phenomenon is the result of the
resonant excitation of surface waves supported by the interfaces of
the left-handed material slab. However, it soon became evident that
this imaging mechanism is strongly sensitive to losses and
dispersion \cite{Garcia_LimitPendry, Smith_LimitPendry, Podolskiy},
besides the fact that practical implementations of left-handed
materials are still a challenge. To overcome these obstacles, other
mechanisms to achieve subwavelength imaging have been studied.
Namely, some researchers explored the fact that photonic crystals
may enable negative refraction close to the band-gap edges
\cite{Luo_PC,Luo_PC_2}. Also, recently a planar magnetoinductive
lens was proposed for near-field imaging at the microwave range
\cite{Marques_magnetoLens}. This approach is based on the excitation
of surface waves at two weakly coupled resonant interfaces to
amplify the evanescent waves.

A sub-diffraction near-field imaging mechanism of a different kind
has been proposed by other researchers \cite{Li_PC, Chien_PC,
Kuo_PC, Zhang_PC, canal, canal_CWM, SWIWM}. This technique is based
neither on negative refraction nor on the amplification of
evanescent waves. The idea is to transport the evanescent waves
through an engineered material (transmission device), rather than to
amplify them. This canalization mechanism is possible if the
incoming wave can couple most of its energy to an electromagnetic
mode of the material with phase velocity nearly independent of the
transverse wave vector (i.e. the associated wave normal surface must
be nearly flat). In our opinion, the best solution proposed thus far
at microwave frequencies is the one reported in \cite{SWIWM}. In
this configuration the engineered material is formed by an array of
perfectly conducting wires (wire medium) that guide the radiation
"pixel by pixel" from the input plane to the output plane. As
discussed in \cite{SWIWM, ResolutionPRE}, the wire medium lens is
capable of transporting subwavelength images without significant
distortion to nearly unlimited distances since the influence of
losses is negligibly small. Moreover, the resolution of the
transmission device is only limited by the spacing between the
wires. The problem with the wire medium lens is that its design
cannot be directly scaled to higher frequencies (infrared and
optical domains) because metals lose their conducting properties,
and consequently the electrodynamics of the equivalent medium is
significantly altered. The straightforward attempts to get
sub-wavelength imaging at optical frequencies using rodded medium
\cite{Ono} encounter problems and limitations related to the effect
of strong losses.

Recently a non-local permittivity model was proposed by the authors
to describe the electrodynamics of plasmonic rods
\cite{Silv_Nonlocalrods}. This study enables the proper dimensioning
and generalization of the transmission device proposed in
\cite{SWIWM} to higher frequencies. In \cite{Silv_Nonlocalrods} it
was proved that a periodic array of rods with negative real part of
permittivity (this is the case of all metals at infrared and optical
frequencies) can be modeled using an effective permittivity tensor
and that spatial dispersion effects need to be considered. Moreover,
our studies showed that despite the rods not being perfectly
conducting, the equivalent medium still supports nearly
dispersionless modes, and we speculated that such modes may be used
to enable near-field subwavelength imaging at the infrared and
optical domains.

In this work, we will extend the theory proposed in
\cite{Silv_Nonlocalrods} and demonstrate that in fact silver
nanorods enable sub-diffraction imaging up to terahertz (THz) and
infrared (IR) region of the electromagnetic spectrum. In recent
years, THz and IR radiation has attracted special attention, because
it offers scientific and technological opportunities in many fields,
such as in telecommunications, in computer technology, in
spectroscopy, and in sensing \cite{SensingTHz}. Near-field imaging
at THz and IR frequencies may be very important for diverse
applications, such as the near-field microscopy of surfaces,
recording information with increased storage capacity, or detecting
and measuring the mass and volume fractions of bio cells (e.g. cells
of blood), many organic molecules, radioactive atoms, microbes,
viruses and explosive matters.

The paper is organized as follows. In section \ref{elecdyn} we
briefly review the electrodynamics of the plasmonic rod medium, and
the possibility of canalizing waves through a rod medium slab. In
section \ref{secdimrods} we explain how to properly design the
electromagnetic crystal in order to ensure that the phase velocity
of waves is independent of the transverse wave vector. Next, in
section \ref{secttx} we introduce an homogenization technique that
enables the accurate analytical characterization of the transmission
properties of a rod medium slab. In section \ref{secdimL} it is
explained how to tune to thickness of the transmission device in
order that it is operated in the canalization regime. In section
\ref{secsims} we present full wave simulations that illustrate the
potentials of the proposed configuration, and demonstrate
subwavelength imaging at 30THz. Finally, in section \ref{secconcl}
the conclusions are presented.

In this work the fields are assumed monochromatic with time
dependence $e^{+ j \omega t}$.

\section{Canalization of waves through the rod medium \label{elecdyn}}

\begin{figure}[h]
\centering \epsfig{file=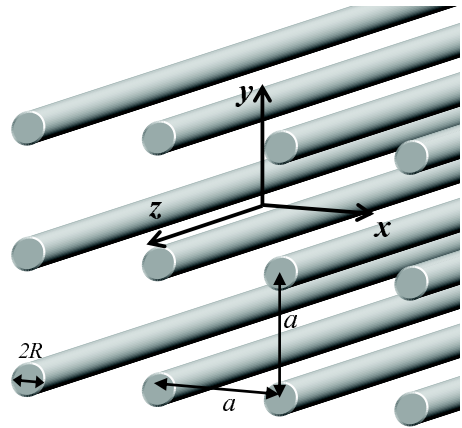, width=6cm} \caption{Periodic
array of long thin plasmonic rods arranged in a square lattice.}
\label{rods}
\end{figure}

In order to enable the proper study and design of waveguiding
structures based on plasmonic rods, next we review the
homogenization theory introduced in \cite{Silv_Nonlocalrods}. Let us
consider a periodic array of plasmonic rods oriented along the
$z$-direction, as depicted Fig. \ref{rods} The rods are infinitely
long, with radius $R$, and the spacing between the rods (lattice
constant) is $a$. The rods are embedded in a dielectric host with
permittivity $\varepsilon _h$. The permittivity of the rods is
$\varepsilon_m$.

In \cite{Silv_Nonlocalrods} it was proved that for long wavelengths
the fields in the periodic structure can be characterized using a
nonlocal permittivity dyadic: \e
\=\varepsilon(\omega,k_z)=\varepsilon_h\left[\varepsilon_{t}(\-x\-x+\-y\-y)+\varepsilon_{zz}(\omega,k_z)\-z\-z\right],\f
\e \varepsilon_{t} =  1 + \frac{2}{{\frac{1}{{f_V
}}\frac{{\varepsilon_m  + \varepsilon _h}}{{\varepsilon_m  -
\varepsilon _h}} - 1}}, \l{epsxx}\f \e \varepsilon _{zz}(\omega,k_z)
= 1 + \frac{1}{{\frac{\varepsilon_h}{{\left( {\varepsilon_m -
\varepsilon_h } \right)f_V }} - \frac{{\beta ^2  - k_z^2 }}{{\beta
_p^2 }}}}, \l{epsfinal} \f where $f_V = {\pi R^2}/a^2 $ is the
volume fraction of the rods, $k_z$ is the $z$-component of the wave
vector, $\beta = \omega \sqrt {\varepsilon _h \mu _0 } $ is the wave
number in the host medium, and $\beta_p$ is the plasma wave number
for perfectly conducting rods given by \cite{WMPRB,
Silv_Nonlocalrods}: \e \left( {\beta _p a} \right)^2 \approx
\frac{{2\pi }}{{\ln \left( {\frac{a}{{2\pi R}}} \right) + 0.5275}}.
\l{betp} \f As discussed in \cite{Silv_Nonlocalrods}, provided the
permittivity of the rods is such that $ |\varepsilon_m| \gg
\varepsilon_h$ and the rods are thin, then $\varepsilon_{t} \approx
1$. That is the case here since the plasma frequency of metals is
well above the far infrared band.

It is well known \cite{OptConst1, OptConst2, OptConst3, MetalPC},
that the permittivity of metals at infrared and optical frequencies
can be accurately described by the Drude model: \e
\varepsilon_m(\omega) = \varepsilon _0 \left(\varepsilon _\infty -
\frac{\omega _m^2 }{\omega^2 - j\omega\Gamma} \right),
 \l{drude} \f where
$\omega_m$ and $\Gamma$ are the plasma and damping frequencies of
the material, respectively. To give further insights into the
permittivity model, let us temporarily consider that losses can be
neglected in the frequency window of interest, and that $\omega \ll
\omega_m$. In these circumstances, substitution of \r{drude} into
\r{epsfinal} gives  \e \varepsilon_{zz} \approx
1-\frac{\beta_p^2}{\left(1+\frac{\beta_p^2}{f_V\beta_m^2}\right)\beta^2-k_z^2},
\l{epssimpl}\f where we put $\beta _m = \omega _m \sqrt {\varepsilon
_0 \mu _0 }$. By direct inspection of \r{epssimpl}, it is clear that
wave propagation in the $xoy$-plane (i.e. $k_z=0$) with electric
field polarized along the rods is only possible when $\beta >
\beta_{\rm cut-off}$, where: \e \beta_{\rm
cut-off}^2=\frac{1}{\frac{1}{\beta_p^2}+\frac{1}{f_V\beta_m^2}}.
\l{cut}\f Hence, the cut-off frequency of the structure is
determined by both the plasmonic properties of the metal and by the
plasmonic properties intrinsic to the geometry of the structure.
\begin{figure}[h]
\centering \epsfig{file=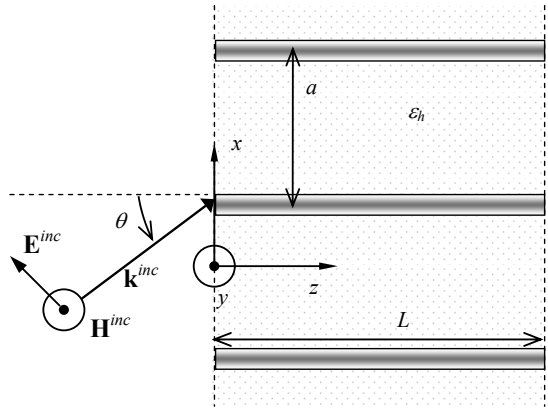, width=8cm} \caption{A plane wave
illuminates a slab of the rod medium (sideview). The thickness of
the slab is $L$ and the rods are normal to the interface. The
structure is periodic along the $x$ and $y$ directions.}
\label{slab}
\end{figure}

The idea to achieve sub-diffraction imaging is to guide the
electromagnetic fields through an array of metallic rods normal to
the front and back interfaces, as illustrated Fig. \ref{slab}. The
canalization phenomenon occurs when the phase velocity of an
electromagnetic mode is nearly independent of the transverse wave
vector. Next, we study in which circumstances can this happen.

We consider that the thickness of the slab is $L$, and that for
$z<0$ and $z>L$ the medium is air. A plane wave with transverse wave
vector ${\bf{k}}_{||}=(k_x,k_y,0)$ illuminates the slab (Fig.
\ref{slab} depicts the case $k_y=0$, which is assumed throughout
this section without loss of generality). The incident magnetic
field is parallel to the interface (the wave is transverse magnetic
with respect to the $z$-direction). As is well-known, the transverse
component of the wave vector is preserved at the boundary. The
incident wave excites two TM-polarized electromagnetic modes in the
rod medium with propagation constants $k_z^{(1)}$ and $k_z^{(2)}$,
respectively: \e \left(k_z^{(1,2)}\right)^2 = \frac{1}{2}\left\{
\varepsilon _{t} \left(\beta ^2 - k_{||}^2 \right) + ( \beta^2 +
\beta _c^2 - \beta _p^2) \right. \l{dispTMkz}\f $$ \left.
 \pm \sqrt {\left[ \varepsilon _{t} \left(\beta ^2-k_{||}^2
\right)  - (  \beta^2 + \beta _c^2 - \beta _p^2) \right]^2  +
4\varepsilon _{t}k_{||}^2\beta _p^2 } \right\}, \nonumber
$$
where \e \beta _c^2  =  - \frac{{\varepsilon_h \beta _p^2 }}{{\left(
{\varepsilon_m - \varepsilon_h} \right)f_V }}. \l{betc}\f Note that
 \r{dispTMkz} was obtained by solving the dispersion equation for
the rod medium in the same manner as in \cite{Silv_Nonlocalrods}; it
simplifies to the result (23) presented in \cite{Silv_Nonlocalrods}
when $\varepsilon _{t} = 1$.

When the permittivity of the rods follows the Drude model \r{drude}
and the losses are negligible, $\beta_c$ is to a first approximation
a positive real number, (with the same unities as $\beta$), that
increases linearly with frequency: \e \beta_c \approx
\frac{\beta_p}{\sqrt{f_V}\beta_m}\beta. \l{betc_apprx}\f

As referred before, the canalization of waves through the engineered
material is possible if either $k_z^{(1)}$ or $k_z^{(2)}$ are nearly
independent of the transverse wave vector ${\bf{k}}_\parallel$
\cite{canal}. In that case all the spatial harmonics suffer the same
phase-shift when they propagate through the rod medium slab (see
Fig. \ref{slab}), provided the length $L$ is also properly chosen
\cite{canal}. In \cite{Silv_Nonlocalrods}, two different
opportunities to get flat isofrequency contours were identified. It
was shown that if either $\beta _c \ll \beta _p$ or $\beta _c \gg
\beta _p$, or equivalently, using \r{betc_apprx} and supposing that
the permittivity of the host is comparable to that of free-space, if
either $\omega\ll \omega_m \sqrt {f_V}$ or $\omega\gg\omega_m \sqrt
{f_V}$, the propagation constant of one of the modes, given by
\r{dispTMkz}, becomes nearly independent of ${\bf{k}}_\parallel$.
The two possibilities are summarized below:
\begin{enumerate}
\item $\omega\ll \omega_m \sqrt {f_V}$. In this case the nearly dispersionless mode is quasi-TEM, and
has characteristics similar to the transmission line modes of the
wire medium \cite{WMPRB}. The radiation is not able to effectively
penetrate into the rods, and so most of the energy propagates in the
host medium. The isofrequency contour of the pertinent mode gets
flatter as $\frac{\beta_p}{\sqrt{f_V}\beta_m}$ decreases. Note that
recently \cite{WG_THZ_Nature, WG_THZ_1}, it was experimentally
proved that a related propagation mechanism enables waveguiding of
electromagnetic energy at THz frequencies.
\item
$\omega \gg \omega_m \sqrt {f_V}$. In this case, the propagation
mechanism is based on the surface plasmon-polariton supported by
each rod. The energy of this surface wave is tightly bounded to the
plasmonic material, and consequently there is no interaction or
coupling between the rods (this is why the mode becomes
dispersionless). When $\frac{\beta_p}{\sqrt{f_V}\beta_m}$ increases
the field confinement also increases, and the isofrequency contours
become more flat.
\end{enumerate}
Which of these possibilities is more interesting to achieve
near-field sub-diffraction imaging? The answer is directly related
to the effect of losses. Indeed, in the first case, since the field
energy is mostly concentrated in the host medium \cite{WG_THZ_1}, it
is expected that the rod material losses have a mild effect on the
propagation properties. Quite differently, in the second case, the
surface plasmon-polariton has a field distribution tightly bounded
to the metallic rods \cite{ENG_WG_Tak}, and so the effect of losses
may be dramatic, and ultimately destroy the possibility of imaging
as it was demonstrated in \cite{Ono}. Hence, in this work we will
explore the first regime, for which, as referred above, the rods
support a quasi-TEM mode.

Firstly, we will study the influence of $\omega_m$ on the properties
of the modes. We consider propagation along the rod axes,
$k_\parallel=0$, and use the fact that $\varepsilon_{t} \approx 1$.
Then \r{dispTMkz} yields,
\begin{eqnarray}
k_{0}^{(1)}\equiv\left.k_z^{(1)}\right|_{k_{||}  = 0} &&= \beta \sqrt{\varepsilon_{t}} \approx \beta \nonumber \\
k_{0}^{(2)}\equiv\left. {k_z^{(2)}} \right|_{k_{||}  = 0} &&\approx
-j\sqrt {\beta _p^2 - \beta _c^2 - \beta ^2 } \l{hTM0}
\end{eqnarray}
Thus, the first mode has no cut-off and propagates nearly with the
speed of light in the dielectric host, while the second mode can
only propagate above the cut-off frequency given by \r{cut}. On the
other hand, letting $k_\parallel\to \infty$, we find that:
\begin{eqnarray}
k_{\infty}^{(1)}\equiv\left. {k_z^{(1)}} \right|_{k_{||} = \infty} &&\approx \sqrt {\beta ^2  + \beta _c^2 } \nonumber \\
k_{\infty}^{(2)}\equiv\left. {k_z^{(2)}} \right|_{k_{||}  = \infty}
&&= -j\infty
\end{eqnarray}
Hence, for off-axis propagation, the attenuation of the second mode
increases. Quite differently, the first mode is never cut-off and
may be nearly dispersionless. Indeed, using \r{betc_apprx} we can
write that, \e k_\infty^{(1)} \approx \beta \sqrt {
1+\frac{\beta_p^2}{f_V\beta_m^2} } \l{hTMinft}\f and so it is
evident that when $\frac{\beta_p}{\sqrt{f_V}\beta_m}$ is small
$k_0^{(1)}$ and $k_\infty^{(1)}$ are nearly identical, and the
isofrequency contour is almost flat. In this case, the mode becomes
quasi-TEM. Notice that if the rods are perfectly conducting then
$k_z^{(1)}=\beta$ for any $k_{||}$ and the mode is completely
dispersionless and exactly TEM (transmission line mode of the wire
medium \cite{WMPRB}). In general, to obtain good imaging at a given
design frequency, the rods material shall be chosen such that the
plasma frequency $\omega_m$ is as high as possible. To give an idea
of the possibilities, we refer that the plasma and damping
frequencies of the following metals are: aluminium (Al) 3570/19.4
THz, copper (Cu) 1914/8.34 THz, gold (Au) 2175/6.5, and silver (Ag)
2175/4.35 THz \cite{OptConst1, OptConst2, MetalPC}.

\section{Dimensioning of the rods \label{secdimrods}}

In order to properly dimension the engineered material, it is also
of obvious importance to assess the effect of the rods spacing $a$,
and radius $R$, on the propagation characteristics.
\begin{figure}[h]
\centering \epsfig{file=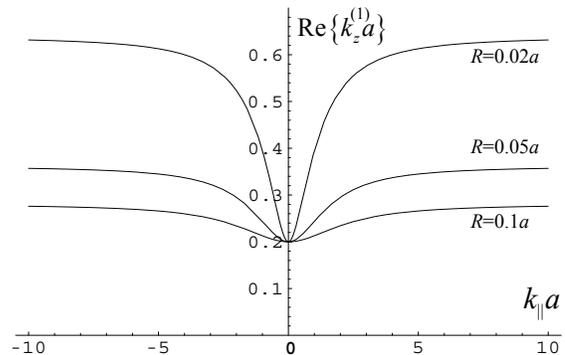, width=8cm} \caption{Dispersion
contours for the quasi-TEM mode at 30 THz for silver rods with
$a=215nm$ and various values of $R$.} \label{isofR}
\end{figure}
In order to study the isofrequency contours of the quasi-TEM mode,
we use \r{betp}, \r{hTM0} and \r{hTMinft} to obtain: \e
\frac{k_\infty^{(1)}}{k_0^{(1)}} \approx \sqrt {1 + \frac{1}{{\left(
{\beta _m R} \right)^2 }}\frac{2}{{\ln \left( {\frac{a}{{2\pi R}}}
\right) + 0.5275}}} \f This result shows that $a$ has little
influence on the dispersion of the quasi-TEM mode, while the radius
$R$ is the main factor that determines how close to unity the ratio
$k_\infty^{(1)}/k_0^{(1)}$ is. Indeed, if the rods are too thin so
that $\beta _m R \ll 1$, the isofrequency contours are not anymore
flat. This is illustrated in Fig. \ref{isofR}, where we plot the
isofrequency contours for a medium of silver rods at 30 THz ($
\varepsilon _{Ag} \approx  - 5143 - 746j $), for different values of
$R$. The host medium is assumed to be Halcogenide glass,
$\varepsilon_h/\varepsilon_0 = 2.2$, and the lattice constant is
$a=215nm$ (i.e. $\beta a =0.2$). In Fig. \ref{isofR} it is seen that
the contours become hyperbolic when the radius of the rods is very
small. Notice that the skin-depth in the rod material is given by
$\delta \approx 1/\beta_m$. Hence, the condition to get flat
contours is equivalent to $\delta \ll R$, i.e. the metal skin-depth
must be much smaller than the radius of the wire.

\begin{figure}[h]
\centering \epsfig{file=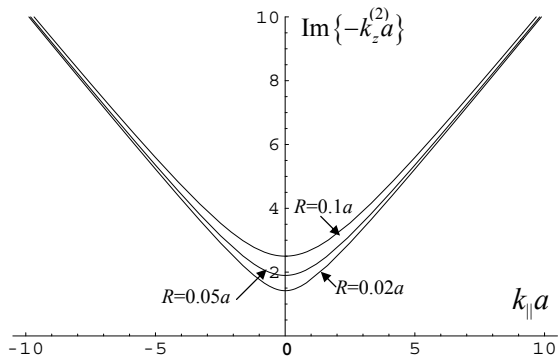, width=8cm} \caption{Normalized
attenuation constant (non-propagating mode) at 30 THz for Silver
rods with $a=215nm$ and different values of $R$.} \label{isoR_gTM}
\end{figure}

For the imaging application it is essential that not only the phase
velocity of one of the modes is independent of $k_\parallel$, but
also that the other mode is severely attenuated so that it cannot be
excited and interfere with the propagating mode. Thus, it is
desirable that $ {\mathop{\rm Im}\nolimits} \{ -k_z^{(2)} \} $ is as
large as possible. From \r{hTM0} one can verify that $ {\mathop{\rm
Im}\nolimits} \{ -k_0^{(2)} a\} $ increases either when the lattice
constant $a$ decreases, or when the ratio $R/a$ increases. As
before, this means that very thin rods may adversely affect the
propagation characteristics of the mode with propagation constant
$k_z^{(2)}$. This is illustrated in Fig. \ref{isoR_gTM}, where we
plot the normalized attenuation constant as function of
$k_\parallel$ for different values of $R$. The rest of the
parameters are as in the previous example. Nevertheless, the crucial
factor that determines how severely the non-propagating mode is
attenuated, is not the radius $R$ but the lattice constant $a$. In
fact, it can be verified that $k_z^{(2)}$ is roughly proportional to
$1/a$. This has a very important implication. Indeed, the total
attenuation (in dBs) for a fixed distance $L$ (see Fig. \ref{slab})
will be proportional to $L/a$. Hence, in general, the smaller the
lattice constant is the more attenuated the mode is. Assuming that
$L$ is comparable to the wavelength in the host material, this
implies that $a$ must be chosen such that $\beta a \ll 1$.

Let us summarize our findings. The spacing between the wires $a$
must be chosen such that $\beta a << 1$. This ensures that the mode
in cut-off is strongly attenuated and does not interfere with the
quasi-TEM mode. On the other hand, to guarantee that the phase
velocity of the quasi-TEM mode is nearly independent of the
transverse wave vector, we need to choose the radius of the rods,
$R$, at least as large as the metal skin-depth $\delta$. Since
$2R<a$, it is clear that the two objectives may be contradictory,
and that a trade-off may be necessary.

Moreover, as discussed next, the ratio $R/a$ must also be kept
relatively small, i.e. large values for $R/a$, let us say $R/a>0.1$,
cannot be admitted (also we remind the reader that the permittivity
model \r{epsfinal} is only valid for diluted systems). The reason
for this restriction is related to the transverse impedance. Indeed,
for thick rods the cross-sectional area occupied by the metal is
large, and this tends to make the transverse impedance small. This
effect is obviously undesired because it deteriorates the matching
with the exterior, and ultimately the bandwidth of the imaging
system.

In practice, when the performance of the engineered material is
mainly limited by the plasmonic properties of the rods, the best
solution is to fix R slightly larger than the metal skin-depth, and
choose $a$ such that $R/a<0.1$. If this solution yields a relatively
small value for $\beta a$, then imaging with good quality can still
be achieved. Notice that for very good conductors, the skin-depth
can be neglected, and so it is always possible to design the
engineered material in such a way that it has excellent imaging
properties \cite{SWIWM, ResolutionPRE}.

\section{Transmission coefficient \label{secttx}}

So far, the possibility of near-field imaging was discussed mostly
in a qualitative way. A more quantitative and detailed study
requires the calculation of the transmission coefficient for the
configuration shown in Fig. \ref{slab}. As referred in
\cite{Silv_Nonlocalrods}, even though the permittivity model of the
medium is known, the calculation of the transmission coefficient is
not trivial. The problem is that a TM-polarized incident wave (i.e.
with magnetic field parallel to the interface) will excite two
distinct TM-modes inside the rod medium slab, as discussed in
section \ref{elecdyn}. This phenomenon is a manifestation of strong
spatial dispersion \cite{Agranovich}, and has an important
consequence: the continuity of the tangential electric and magnetic
fields at the interface is insufficient to calculate the scattering
parameters; note that for TM-incidence, the TE-mode, with electric
field parallel to the interface, cannot be excited
\cite{Silv_Nonlocalrods}. Thus, because of the additional TM-mode in
the artificial material slab, the system has one degree of freedom.
To remove this degree of freedom, an additional boundary condition
(ABC), besides the continuity of the tangential electromagnetic
fields, is necessary.

In \cite{MarioABC} an ABC was formulated and successfully validated
for the case of perfectly conducting rods. The ABC establishes that,
\e\varepsilon _h E_n \;{\rm{is}}\;{\rm{continuous}} \l{ABC}\f at the
interfaces, where $E_n$ is the normal component of the (average)
electric field (i.e. $E_z$ in Fig.\ref{slab}) and $\varepsilon _h$
is the permittivity of the host medium (which in general can be
different in the rod medium slab and in the regions $z<0$ and
$z>L$). Can such ABC be used for the case of non perfectly
conducting rods?

In what follows, we demonstrate that the answer is positive. We will
show  that the physical arguments used in \cite{MarioABC} can be
adapted to the case of plasmonic rods. Let us consider the structure
depicted in Fig.\ref{slab}. The polarization current (relative to
the host medium) inside the plasmonic rods is given by ${\bf{J}}_d =
j\omega \left( {\varepsilon_m - \varepsilon _h } \right){\bf{E}}$,
where $\varepsilon_m$ is the permittivity of the rods (for perfectly
conducting rods, the polarization current flows uniquely on the
surface of the rods). At the interfaces the path along which the
current flows (i.e. the rods) has an abrupt discontinuity. Indeed,
provided $|\varepsilon_m/\varepsilon_0|\gg 1$ the polarization
current lines are not able to penetrate into the regions $z<0$ and
$z>L$, and thus the $z$-component of the polarization current must
be nearly null at the interfaces. Moreover, as in \cite{MarioABC},
it can be proved that following relation holds: \e
\left(\frac{\partial }{{\partial z}}{\bf{\hat u}}_z  -
j{\bf{k}}_{||}  \right) \times {\bf{H}}_{{\rm{av}}} = j\omega
\varepsilon _h(z) {\bf{E}}_{{\rm{av}}} + \frac{1}{a^2}
 {\int\limits_{D} {{\bf{J}}_d e^{j{\bf{k}}_{||} {\bf{.r}}} dx{\kern 1pt} dy  } }
\f where $ {\bf{E}}_{\rm{av}}\left( {z;{\bf{k}}_{||} } \right)$ and
${\bf{H}}_{\rm{av}}\left( {z;{\bf{k}}_{||} } \right)$ represent the
fields averaged over a $z=const.$ cross-section of the unit cell,
and $D$ is the cross section of the rod. But since the tangential
average electromagnetic fields are continuous at the interfaces and,
as discussed above, the $z$-component of the polarization current
vanishes at the interfaces, the above relation implies that the ABC
\r{ABC} holds, as we wanted to establish (for more details see
\cite{MarioABC}).

Using \r{ABC}, we can now easily study the transmission properties
of the rod medium slab. Referring to Fig. \ref{slab}, and assuming
that the incident magnetic field is along the $y$-direction and has
amplitude $H_{0}$, we can write that:

\e H_y = H_{0} e^{ - jk_x x} \left\{
\begin{array}{lcl}
e^{ - j k_z^{(a)} z}  + \rho \,e^{j k_z^{(a)} z}, && z < 0 \\
\hspace{-1.5mm}\begin{array}{lcl}
A_1 e^{ -j k_z^{(1)}z}  + A_2 e^{j k_z^{(1)} z}\\
+ B_1 e^{ -j k_z^{(2)}z} + B_2 e^{j k_z^{(2)}z},\\
\end{array} \hspace{-2mm} &&0 < z < L \\
T\,e^{ - j k_z^{(a)}\left( {z - L} \right)}, &&z >
L\\
\end{array}
\right. \f 
where $\rho$ is the reflection coefficient, $T$ is the transmission
coefficient, $ k_z^{(a)} =-j \sqrt {k_{\parallel}^2 - \varepsilon_0
\mu_0 \omega ^2}$ is the longitudinal component of the wave vector
in the free space regions, and $A_1$, $A_2$, $B_1$, and $B_2$ are
the unknown amplitudes of the TM-modes in the slab. We remember that
for the case illustrated in Fig. \ref{slab}, the transverse wave
vector is ${\bf{k}}_{||}  = \left( {k_x ,0,0} \right)$. For a
propagating incident wave, the angle of incidence is such that $\sin
\theta = k_{\parallel} / \beta_0$ where $\beta_0 = \omega \sqrt
{\varepsilon _0 \mu _0 }$ is the free-space wave number.

It can be proven that the continuity of the tangential
electromagnetic fields and the ABC are equivalent to
\cite{MarioABC}:
\begin{eqnarray}
\left[ {H_y } \right] = 0,\quad{\rm{  }}\quad \left[
{\frac{1}{{\varepsilon _h(z) \varepsilon _{t}\left(z\right) }}\frac{{dH_y }}{{dz}}} \right] = 0, \\
\left[ {\frac{1}{{\varepsilon _{t}(z) }}\frac{{d^2 H_y }}{{dz^2 }} +
\omega ^2 \mu _0 \varepsilon _h(z) H_y } \right] = 0. \nonumber
\end{eqnarray}
In the above the rectangular brackets $[...]$ represent the jump
discontinuity of a given quantity at the interface (i.e. the
quantity evaluated at the air side subtracted from the quantity
evaluated at the rod medium side), and, by definition,
$\varepsilon_h = \varepsilon_0$ and $\varepsilon_{t} = 1$ at the
free space side. Using these boundary conditions, we readily obtain
the following linear system of equations,
\begin{widetext}
\e
\left( {\begin{array}{*{20}c}
   { - 1} & 1 & 1 & 1 & 1 & 0  \\
   { - \frac{k_z^{(a)}}{\varepsilon_{0}}} & { - \frac{k_z^{(1)}}{\varepsilon_{h}\varepsilon_{t}} } & { \frac{k_z^{(1)}}{\varepsilon_{h}\varepsilon_{t}}} & { - \frac{k_z^{(2)}}{\varepsilon_{h}\varepsilon_{t}} } & { \frac{k_z^{(2)}}{\varepsilon_{h}\varepsilon_{t}} } & 0  \\
   -(k_z^{(h)})^2 & \frac{(k_z^{(1)})^2}{\varepsilon_{t} } & { \frac{(k_z^{(1)})^2}{\varepsilon_{t}} } & { \frac{(k_z^{(2)})^2}{\varepsilon_{t}} } & { \frac{(k_z^{(2)})^2}{\varepsilon_{t}} } & 0  \\
   0 & {e^{ - jk_z^{(1)} L} } & {e^{ + jk_z^{(1)} L} } & {e^{ - jk_z^{(2)} L} } & {e^{ + jk_z^{(2)} L} } & { - 1}  \\
   0 & { - \frac{k_z^{(1)}}{\varepsilon_{h}\varepsilon_{t}} e^{ - jk_z^{(1)} L} } & { \frac{k_z^{(1)}}{\varepsilon_{h}\varepsilon_{t}} e^{ + jk_z^{(1)} L} } & { - \frac{k_z^{(2)}}{\varepsilon_{h}\varepsilon_{t}} e^{ - jk_z^{(2)} L} } & {\frac{k_z^{(2)}}{\varepsilon_{h}\varepsilon_{t}} e^{ + jk_z^{(2)} L} } & \frac{k_z^{(a)}}{\varepsilon_{0}}  \\
   0 & {\frac{(k_z^{(1)})^2}{\varepsilon_{t}} e^{ - j k_z^{(1)} L} } & {\frac{(k_z^{(1)})^2}{\varepsilon_{t}} e^{ + j k_z^{(1)} L} } & {\frac{(k_z^{(2)})^2}{\varepsilon_{t}} e^{ - j k_z^{(2)} L} } & { \frac{(k_z^{(2)})^2}{\varepsilon_{t}} e^{ + j k_z^{(2)} L} } & -(k_z^{(h)})^2  \\
\end{array}} \right)\left( {\begin{array}{*{20}c}
   \rho  \\
   {A_1 }  \\
   {A_2 }  \\
   {B_1 }  \\
   {B_2 }  \\
   T  \\
\end{array}} \right) = \left( {\begin{array}{*{20}c}
   1  \\
   - \frac{k_z^{(a)}}{\varepsilon_{0}}\\
   (k_z^{(h)})^2  \\
   0  \\
   0  \\
   0  \\
\end{array}} \right) \l{linearsys}\f
where we defined $ k_z^{(h)} =-j \sqrt {k_{\parallel}^2 -
\varepsilon_h \mu_0 \omega ^2}$. In general, the solution of system
\r{linearsys} must be obtained numerically. In the particular case
where the rods are embedded in air ($\varepsilon_h=\varepsilon_0$),
and within the approximation $\varepsilon_{t}=1$, the transmission
coefficient $T$ can be written in a compact closed analytical form
(an analogous result was obtained in \cite{ResolutionPRE} for the
case of perfectly conducting rods):
\begin{eqnarray}
T = \frac{1}{{1 + \frac{{jk_z^{\left( 1 \right)} }}{{k_z^{\left( a
\right)} }}\tan \left( {\frac{{k_z^{\left( 1 \right)} L}}{2}}
\right)\frac{{\left( {k_z^{\left( a \right)} } \right)^2  - \left(
{k_z^{\left( 2 \right)} } \right)^2 }}{{\left( {k_z^{\left( 1
\right)} } \right)^2  - \left( {k_z^{\left( 2 \right)} }
\right)^2}}+ \frac{{jk_z^{\left( 2 \right)} }}{{k_z^{\left( a
\right)} }}\tan \left( {\frac{{k_z^{\left( 2 \right)} L}}{2}}
\right)\frac{{\left( {k_z^{\left( a \right)} } \right)^2  - \left(
{k_z^{\left( 1 \right)} } \right)^2 }}{{\left( {k_z^{\left( 2
\right)} } \right)^2 - \left( {k_z^{\left( 1 \right)} } \right)^2 }}
}} \nonumber \\
- \frac{1}{{1 - \frac{{jk_z^{\left( 1 \right)} }}{{k_z^{\left( a
\right)} }}\cot \left( {\frac{{k_z^{\left( 1 \right)} L}}{2}}
\right)\frac{{\left( {k_z^{\left( a \right)} } \right)^2  - \left(
{k_z^{\left( 2 \right)} } \right)^2 }}{{\left( {k_z^{\left( 1
\right)} } \right)^2  - \left( {k_z^{\left( 2 \right)} } \right)^2
}}- \frac{{jk_z^{\left( 2 \right)} }}{{k_z^{\left( a \right)} }}\cot
\left( {\frac{{k_z^{\left( 2 \right)} L}}{2}} \right)\frac{{\left(
{k_z^{\left( a \right)} } \right)^2  - \left( {k_z^{\left( 1
\right)} } \right)^2 }}{{\left( {k_z^{\left( 2 \right)} } \right)^2
- \left( {k_z^{\left( 1 \right)} } \right)^2 }} }}  \l{tx}
\end{eqnarray}
\end{widetext}

In order to validate the proposed ABC and the derived formulas, next
we compare the results yielded by our homogenization theory against
full wave numerical simulations \cite{CST} . We consider that the
rods are made of silver, as in section \ref{secdimrods}. The
frequency is 30 THz, the lattice constant is $215$nm ($\beta a =
0.2$), the radius is $R=21.5$nm ($R=0.1 a$), and the length of the
rods is $L= 5.93 \mu$m ($\beta L = 1.76 \pi$). As before, the rods
are embedded in a host medium with $\varepsilon_h/\varepsilon_0 =
2.2$.
\begin{figure}[h]
\centering \epsfig{file=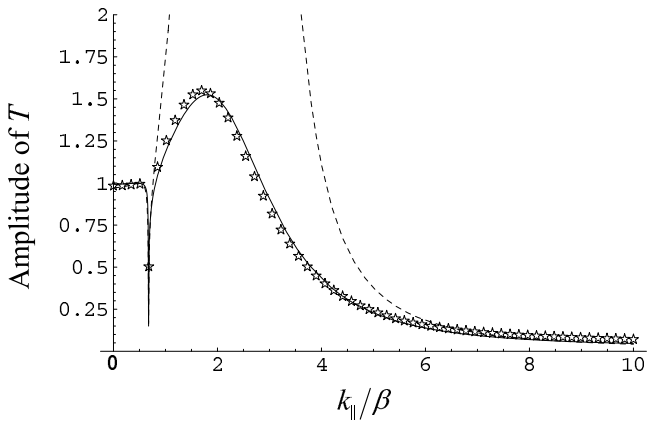, width=8.5cm} \caption{Amplitude
of the transmission coefficient as a function of $k_{\parallel} /
\beta$ for silver rods at 30 THz. The "stars" represent the full
wave result, the dashed line represents the analytical result, and
the solid line corresponds to the analytical result with corrected
$k_z^{(1)}$.} \label{txAbsEx1}
\end{figure}
\begin{figure}[h]
\centering \epsfig{file=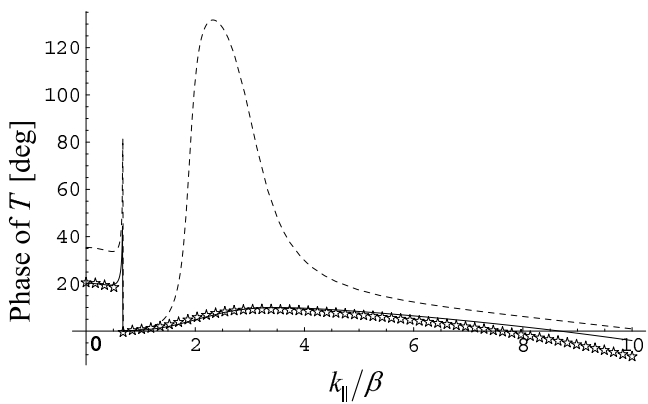, width=8.5cm} \caption{Phase of
the transmission coefficient as a function of $k_{\parallel} /
\beta$ for silver rods at 30 THz. The legend is as in figure
\ref{txAbsEx1}.} \label{txArgEx1}
\end{figure}

In Fig. \ref{txAbsEx1} and Fig. \ref{txArgEx1} we depict the
amplitude and phase, respectively, of the transmission coefficient
as a function of $k_{\parallel} / \beta$. The star symbols represent
the full wave simulations, whereas the dashed line represents the
analytical results. Note that for evanescent waves the transmission
coefficient may be larger than unity without violating the
conservation of energy. As seen, the agreement is not very
satisfactory, particularly for the evanescent spatial harmonics
($k_{||}>\beta_0$). After extensive numerical simulations, we found
out that apparently the justification is that the analytical model
\r{epsfinal} may slightly underestimate the propagation constant
$k_z^{(1)}$. This means that the propagation velocity along $z$ is
slightly less than the velocity of light in the host medium. This
second order effect may be caused by the effect of higher order
multipoles. A simple way to obtain more accurate results is to
multiply $k_z^{(1)}$ by a correction factor. The correction factor
depends, of course, on the geometry of the crystal and on the
permittivity of the rods (for perfectly conducting rods no
correction factor is required). For the set of parameters of this
example, we verified that the correction factor is $1.04$ (around 30
THz). The transmission coefficient calculated with the corrected
$k_z^{(1)}$ corresponds to the solid line in Fig. \ref{txAbsEx1} and
Fig. \ref{txArgEx1}. As seen, the agreement becomes very good.
Similar results are obtained for the reflection coefficient and
other geometries. We stress that the correction factor is
independent of $L$, and thus the analytical model is very convenient
to make parametric studies and understand the electromagnetic
response of the rod medium slab.

\section{Dimensioning of the slab thickness \label{secdimL}}

In this section, we will study the imaging properties of the system
depicted in Fig. \ref{slab}, and discuss how the length $L$ of the
rods is chosen. As is well-known, a wave at the front plane, $z=0$,
of the imaging system can be decomposed into spatial harmonics. Each
Fourier harmonic is associated with a different transverse wave
vector $\bf{k}_{||}$. Because of the superimposition principle, the
Fourier harmonics travel independently through rod medium slab.
Thus, the transmission coefficient $T$ can be regarded as a transfer
function. For the ideal reconstruction of the image at the output
plane $z=L$, the amplitude and phase of $T$ should be ideally
independent of $k_{||}$ (we note here parenthetically that for our
system this may be possible only for the spatial harmonics with
TM-$z$ polarization; the TE-$z$ Fourier harmonics are not guided by
the metallic rods).

In order to make $T$ independent of $k_{||}$, it was suggested in
\cite{canal} to operate the transmission device under a Fabry-Perot
resonance. More specifically, one should choose $L$ in such a way
that: \e k_z^{(1)} L = n \pi, \quad n=1,2,..\l{Fabry} \f Note that
the integer $n$ determines the thickness of the system. Provided
single mode operation can be assumed (i.e. provided the effect of
the evanescent TM-mode can be ignored) it is straightforward to
prove that the transmission coefficient becomes $T = \mp 1$, where
the "$-$" sign is chosen for $n$ odd, and "$+$" sign for $n$ even.
The remarkable result \cite{canal} is that if $k_z^{(1)}$ is
independent of $k_{||}$, the Fabry-Perot condition is fulfilled
simultaneously for all the spatial harmonics, and so the imaging
system is perfect. In practice, the resolution of the system will be
limited by the evanescent TM-mode, and also by the dispersion in the
phase velocity of the quasi-TEM mode.

First we analyze the effect of the evanescent mode. For simplicity,
we will assume that the rods are embedded in air so that the
analytical formula \r{tx} can be used. Taking into account the
Fabry-Perot resonance condition \r{Fabry}, and assuming that the
evanescent TM-mode is very much attenuated across the slab,
${\mathop{\rm Im}\nolimits} \left\{ {-k_z^{(2)} L} \right\} \gg
1$,we obtain that: \e T \approx  \mp \frac{1}{{1 +
\frac{{k_z^{\left( 2 \right)} }}{{k_z^{\left( a \right)}
}}\frac{{\left( {k_z^{\left( a \right)} } \right)^2  - \left(
{k_z^{\left( 1 \right)} } \right)^2 }}{{\left( {k_z^{\left( 2
\right)} } \right)^2  - \left( {k_z^{\left( 1 \right)} } \right)^2
}}}} \l{txap}\f where the "$-$" and "$+$" signs are chosen as
before. Since $k_z^{(a)}$ and $k_z^{(1,2)}$ are in general functions
of $k_{\parallel}$, the above expression also is, and the imaging is
not perfect, even in the case of perfectly conducting rods where
$k_z^{(1)}$ is exactly dispersionless \cite{ResolutionPRE}. When the
attenuation constant of the evanescent mode is very large,
$|k_z^{(2)}| \gg |k_z^{(a)}|$ and $|k_z^{(2)}| \gg |k_z^{(1)}|$,
equation \r{txap} yields $T \approx \mp 1$, i.e. the result that we
got when assuming single mode operation.

Let us discuss now the effect of non-flat isofrequency contours for
the quasi-TEM mode, i.e. the consequences of $k_z^{(1)}$ being a
function of $k_{\parallel}$. This non-ideal behavior means that
there is no collective Fabry-Perot resonance: the electrical length
corresponding to the resonance becomes a function of
$k_{\parallel}$. Thus, even if the effect of the evanescent TM-mode
could be neglected, the imaging would not be perfect. Indeed, from
section \ref{elecdyn} we know that $k_z^{(1)}$ has the following
bounds: \e k_0^{(1)}  \le k_z^{(1)} \left( {k_{||} } \right) \le
k^{(1)}_\infty \f Hence, to recover a certain spectral component of
the image at the output plane, $L$ must be chosen such that it
satisfies \r{Fabry} for the pertinent $k_{||}$, and consequently the
optimum $L$ becomes a function of $k_{||}$: \e L_\infty \le L \le
L_0 \l{length} \f In above, $L_{0,\infty} = {\mathop{\rm
Re}\nolimits} \left\{ { n \pi/k_{0,\infty}^{(1)} } \right\}$. Note
that $L_0$ can be written as $L_0 = \frac{ n \lambda_h}{2
\varepsilon_{t}}$, where $\lambda_h$ is the wavelength in the host
material.

\begin{figure}[h]
\centering \epsfig{file=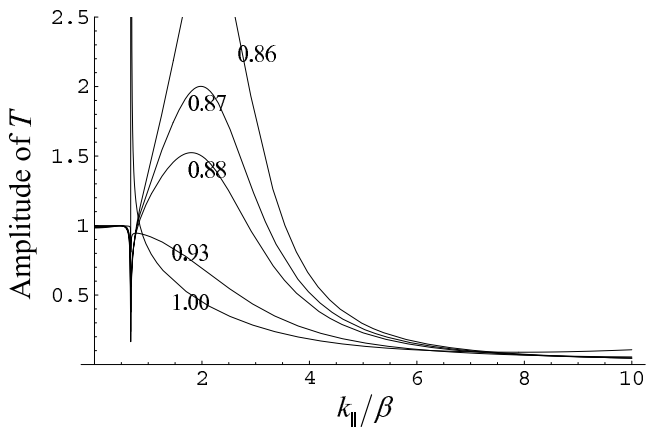, width=8.5cm} \caption{Amplitude of
the transmission coefficient as a function of $k_{\parallel} /
\beta$ for silver rods at 30 THz. The length of the rods is $L = q
\lambda_h$, where $\lambda_h = 6.74 \mu$m and $q = 1.00, 0.93, 0.88,
0.87, 0.86$ (each value of $q$ is associated with a different
curve).} \label{txAbsStudy}
\end{figure}
\begin{figure}[h]
\centering \epsfig{file=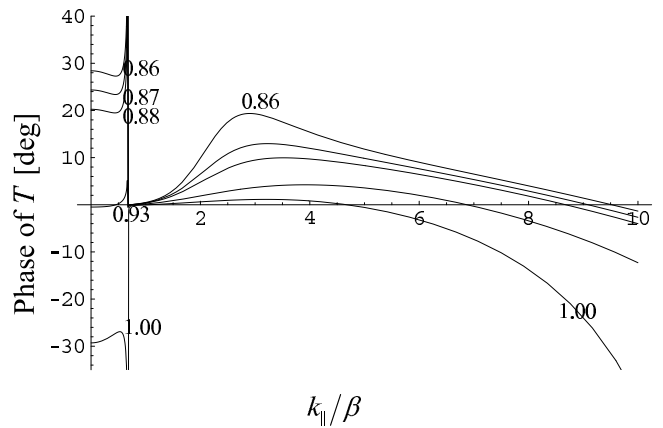, width=8.5cm} \caption{Phase of the
transmission coefficient as a function of $k_{\parallel} / \beta$
for silver rods at 30 THz. The length of the rods is $L = q
\lambda_h$, where $\lambda_h = 6.74 \mu$m and $q = 1.00, 0.93, 0.88,
0.87, 0.86$.} \label{txArgStudy}
\end{figure}
\begin{figure}[h]
\centering \epsfig{file=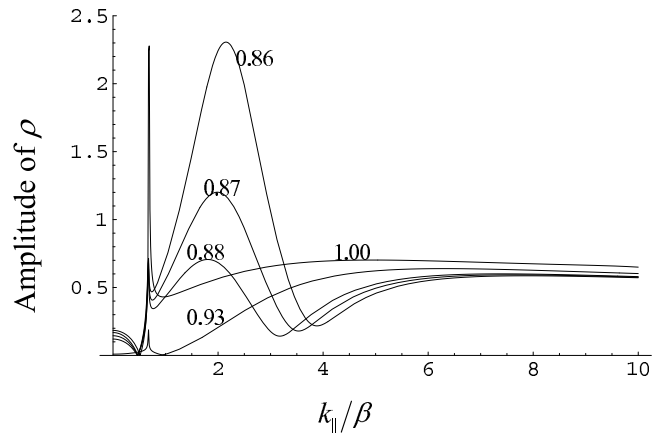, width=8.5cm} \caption{Amplitude of
the reflection coefficient as a function of $k_{\parallel} / \beta$
for silver rods at 30 THz. The length of the rods is $L = q
\lambda_h$, where $\lambda_h = 6.74 \mu$m and $q = 1.00, 0.93, 0.88,
0.87, 0.86$.} \label{roAbsStudy}
\end{figure}
In practice, how should one choose $L$? If we want to recuperate
only the propagating waves, the choice $L=L_0$ is adequate. However,
in general, the condition $L=L_0$ is not appropriate for the
evanescent harmonics with $k_\parallel > \beta_0$. For these modes
the Fabry-Perot condition is fulfilled for some length
$L(k_\parallel)< L_0$. Thus, in order that near-field imaging is
possible, we may have to make a trade-off. More specifically, we may
have to allow some phase and/or amplitude distortion for the
propagating waves, in order to improve transmission of the
evanescent waves. Thus, the length $L$ will be chosen in the
interval defined by \r{length}.

To illustrate the discussion let us consider an example. Again, we
consider silver rods at 30 THz, with the same lattice constant, host
medium, and radius as in the previous section. In Fig.
\ref{txAbsStudy} and Fig. \ref{txArgStudy} the amplitude and phase
of the transmission coefficient are plotted for different values of
$L$. We used the analytical model \r{tx} to plot the curves, taking
also into account the correction factor $1.04$ to better predict
$k_z^{(1)}$, as discussed in the previous section. It is seen that
when the thickness of the rods is close to $0.93\lambda_h$ (which
corresponds to the $n=2$ Fabry-Perot resonance with $L=L_0$, using
the correction factor $1.04$), the transmission coefficient is $T
\approx 1$ for the propagating waves. However, the amplitude
response of the evanescent waves is relatively poor. As discussed
before, the response of the evanescent spectrum can be improved by
decreasing $L$. It is seen that when this is done the phase of $T$
for the propagating waves becomes progressively more distant from
zero, even though it remains relatively flat. However, the amplitude
response of the evanescent spectrum is significantly improved, and
for $L=0.88\lambda_h = 5.93 \mu$m, near field imaging seems possible
roughly for $k_{\parallel}/\beta < 3.5$ with relatively small phase
and amplitude distortion. Note, that if $L$ is greater than $L_0$
(curve with $L = 1.00\lambda_h$) then the transmission
characteristic has a pole very close to $k_{\parallel}/\beta \approx
1$. This indicates that for this set of parameters a surface wave is
supported by the rod medium slab. This regime must be avoided since
the excitation of surface waves and the selective amplification of
very specific spatial harmonics leads to destruction of
subwavelength imaging \cite{ResolutionPRE}.

Another important aspect is to avoid interference between the
reflected wave and the source. To this end, we need to ensure that
the reflection coefficient amplitude remains relatively small in the
frequency band of interest. In Fig. \ref{roAbsStudy} it is seen that
reflection level is very high for $L\le0.87\lambda_h$, and so the
length of the rods must definitely be chosen larger than
$0.87\lambda_h$.

\begin{figure}[h]
\centering \epsfig{file=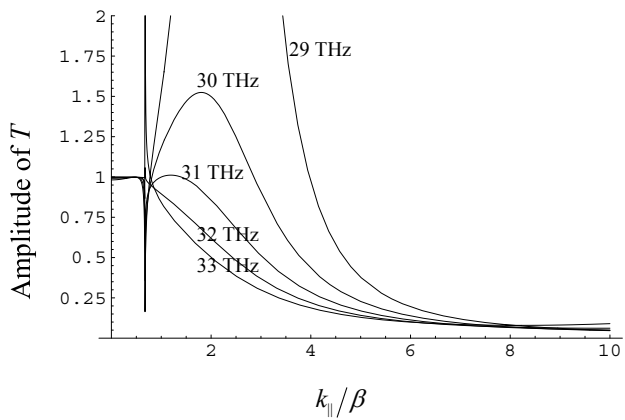, width=8.5cm}
\caption{Amplitude of the transmission coefficient as a function of
$k_{\parallel} / \beta$ for silver rods at different frequencies.
The length of the rods is $L = 5.93 \mu$m.} \label{frqStudy}
\end{figure}

To conclude this section, we study the bandwidth of the proposed
transmission device. In Fig. \ref{frqStudy} the transmission
characteristic is depicted for different frequencies assuming that
the length of the rods is $L = 5.93 \mu$m (corresponds to $q=0.88$
at 30THz). It is seen that the effect of changing frequency for a
fixed $L$ is qualitatively similar to the effect of changing $L$ for
a fixed frequency (Fig.~\ref{txAbsStudy}).

\section{Near-field imaging at infrared frequencies \label{secsims}}

\begin{figure}[h]
\centering \epsfig{file=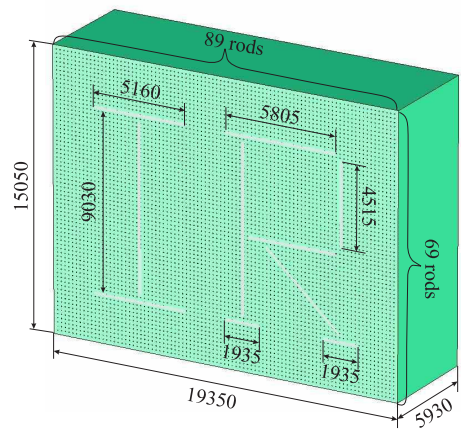, width=6.0cm} \caption{Geometry of
finite-sized slab of rod medium used in full-wave simulation. The
critical parameters are given in~nm.} \label{geomnum} \end{figure}

Next, we illustrate the possibilities and potentials of the proposed
subwavelength imaging mechanism. To this end, we simulated the
performance of a rod medium slab that consists of an array of
$89\times 69$ silver nanorods (see Fig. \ref{geomnum})
using CST Microwave Studio full-wave electromagnetic simulator
\cite{CST}. The nanorods are embedded in Halcogenide glass
($\varepsilon_h=2.2\varepsilon_0$) with total dimensions of $19.35
\times 15.05 \times 5.93 \mu$m. The lattice constant is $a=215$nm,
the radius of the rods is $R=21.5$nm, and the length of the rods is
$L= 5.93 \mu$m, consistently with the simulations presented in the
previous sections. Namely, at 30THz the dispersion contours of the
electromagnetic modes are presented in Fig. \ref{isofR} and Fig.
\ref{isoR_gTM} (curve with $R=0.1a$) and the transmission
coefficient is shown in Fig. \ref{frqStudy} for different
frequencies. We stress that in all the simulations presented in this
paper the effect of realistic losses in silver is taken into
account.

\begin{figure}[h]
\centering \epsfig{file=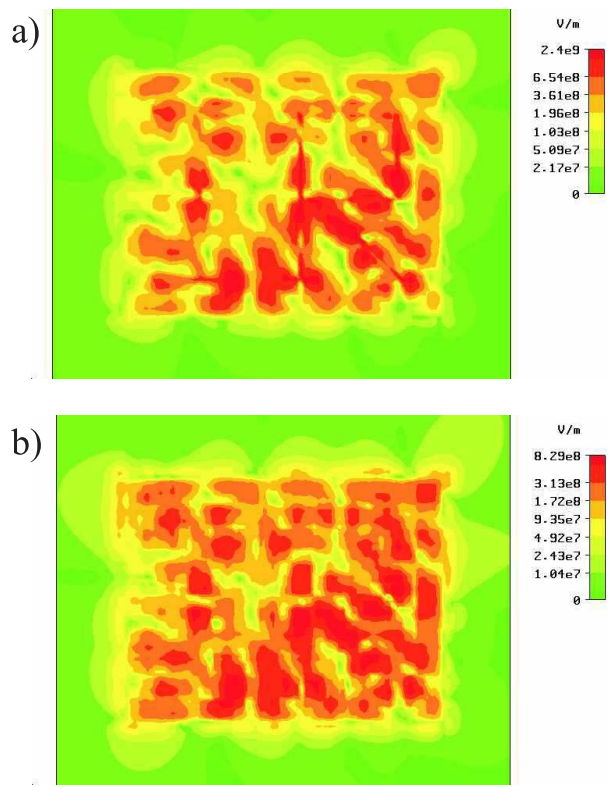, width=8cm} \caption{Distributions of
$|E_z|$ at 29THz: a) at the source plane, at distance $a/2$ from the
source and at the distance $a$ from the front interface, and b) at
the image plane, at distance $a/2$ from the back interface.}
\label{29}
\end{figure}
\begin{figure}[t]
\centering \epsfig{file=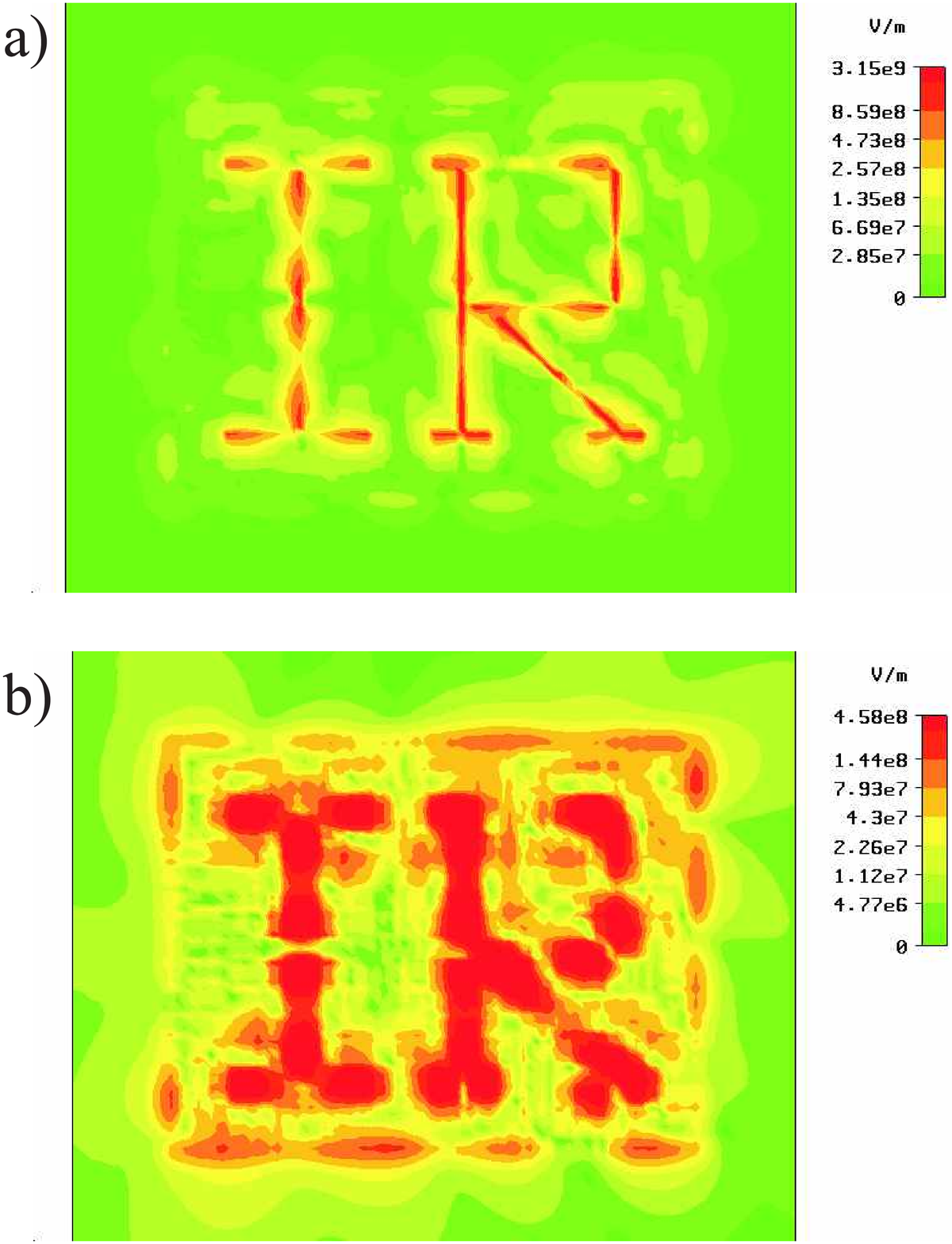, width=8cm} \caption{Distributions of
$|E_z|$ at 30THz: a) at the source plane, and b) at the image
plane.} \label{30}
\end{figure}
\begin{figure}[t]
\centering \epsfig{file=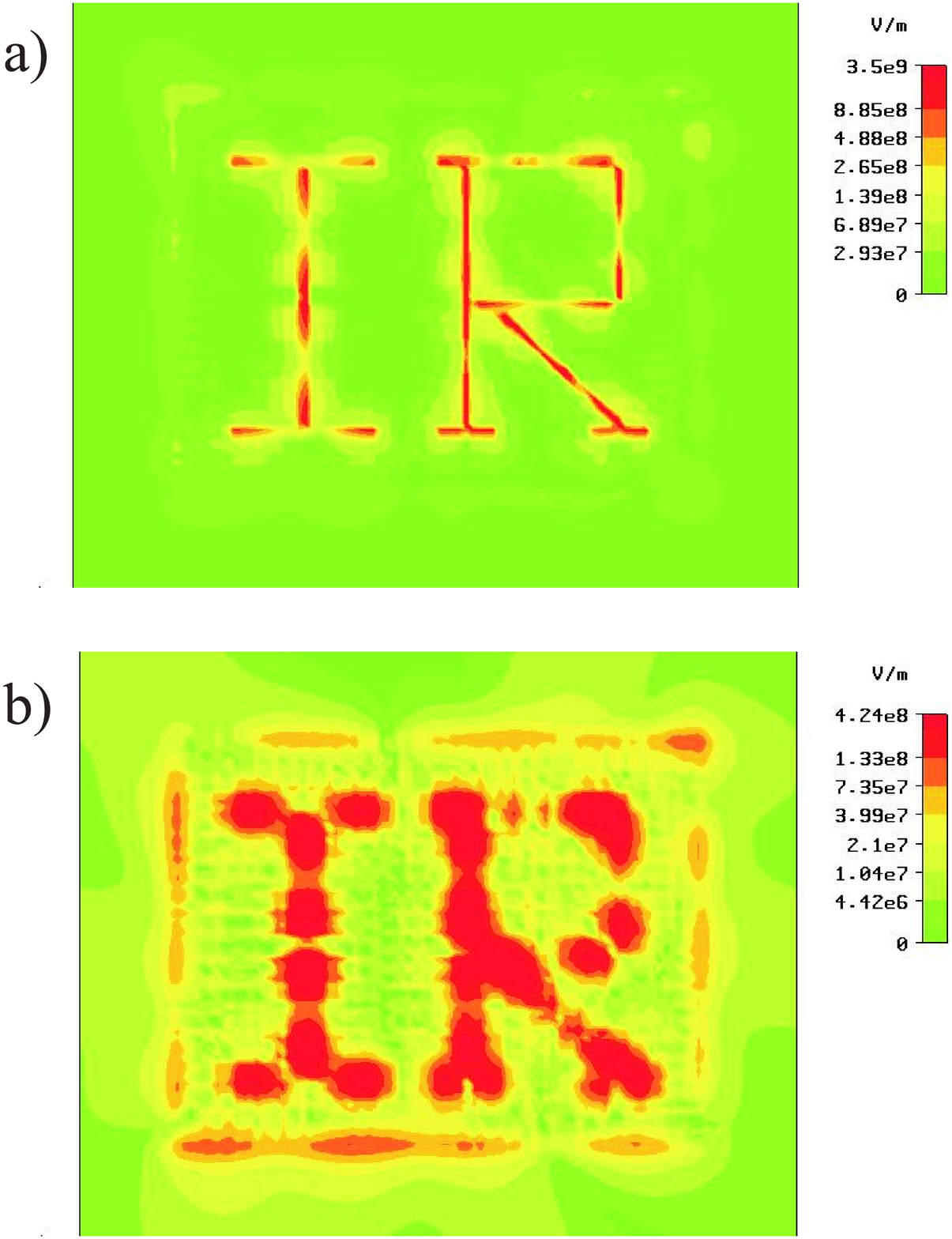, width=8cm} \caption{Distributions of
$|E_z|$ at 31THz: a) at the source plane, and b) at the image
plane.} \label{31}
\end{figure}
\begin{figure}[t]
\centering \epsfig{file=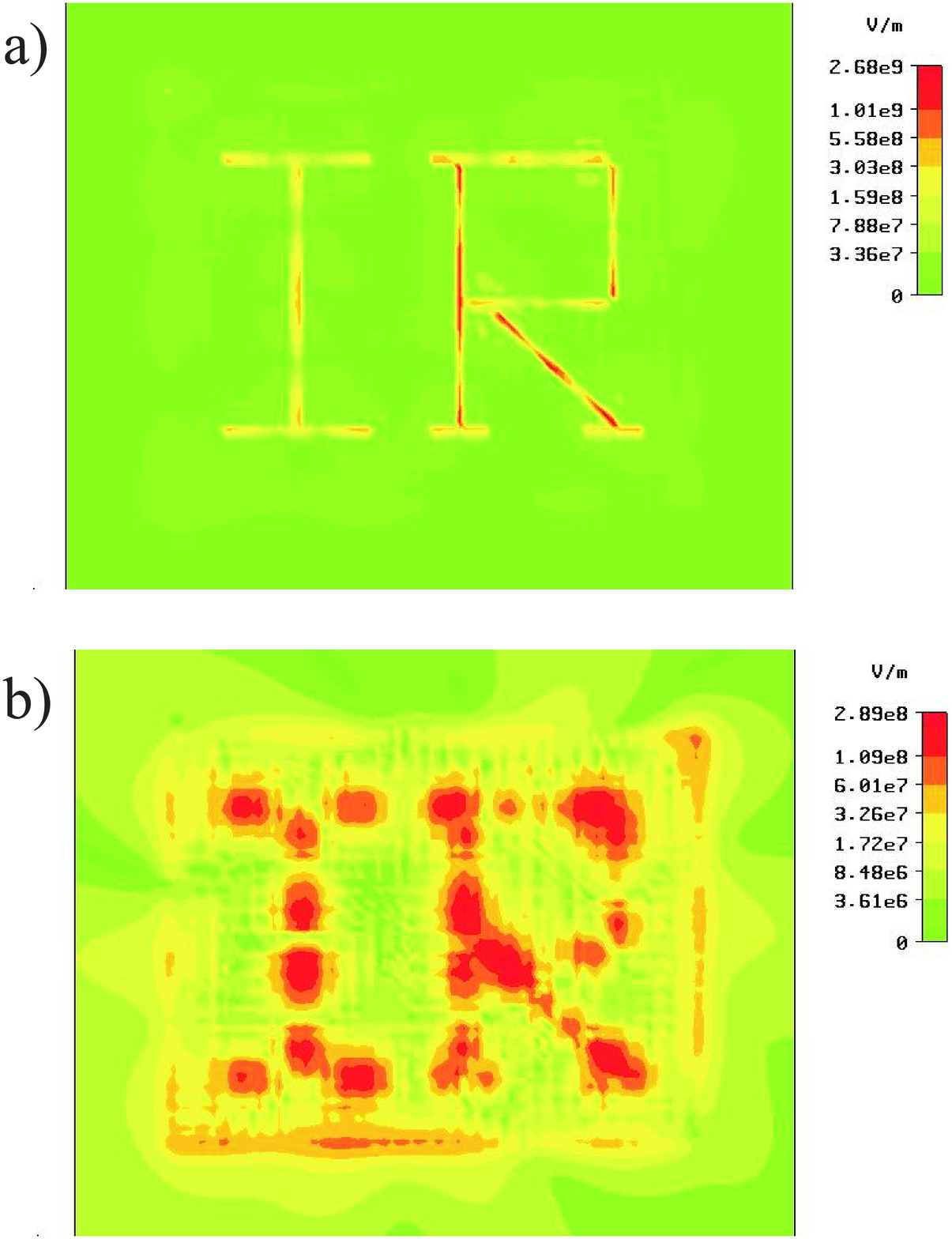, width=8cm} \caption{Distributions of
$|E_z|$ at 33THz: a) at the source plane, and b) at the image
plane.} \label{33}
\end{figure}
\begin{figure}[t]
\centering \epsfig{file=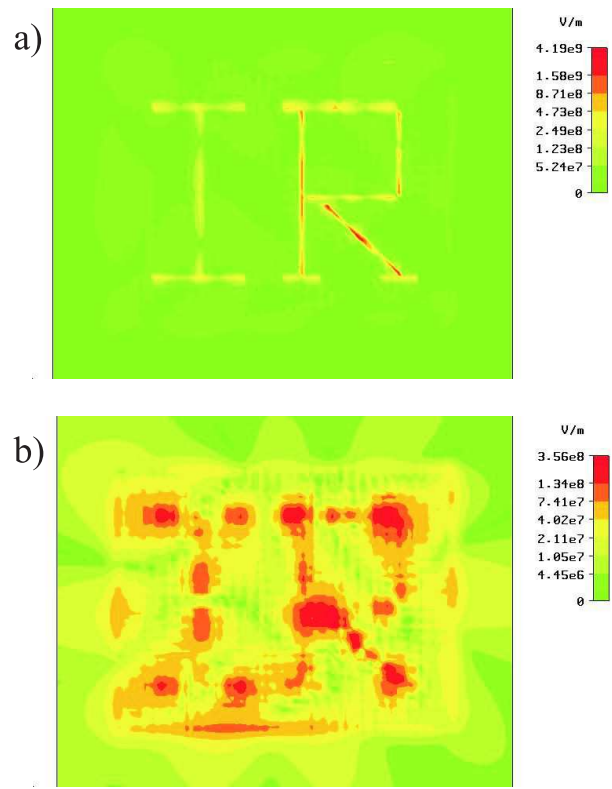, width=8cm} \caption{Distributions of
$|E_z|$ at 36THz: a) at the source plane, and b) at the image
plane.} \label{36}
\end{figure}

In our numerical model the source consists of a set of perfectly
conducting wire dipoles driven by $1A$ currents in their centers and
arranged in the form of letters "IR", see Fig. \ref{geomnum}. The
source is placed at the distance $a/2=107.5$nm from the front
interface. The "IR"-shaped dipole source radiates both TE-$z$ and
TM-$z$ waves. As discussed before, the rod medium slab is able to
guide only the TM-$z$ spatial harmonics. Thus, in order to evaluate
the waveguiding properties of the rod medium slab we numerically
calculated the $z-$component of the electric field at the front
(input) plane and at the back (output) plane. More specifically, the
field is calculated at the distance $a/2$ from the source and back
interface, respectively. Note, that the $z$-component of electrical
field includes only the contribution of  TM-$z$ waves since $E_z =
0$ for the TE-$z$ waves.

In Fig. \ref{29} $|E_z|$ is depicted at 29THz at the input and
output planes, respectively. It is clear that the imaging properties
are very poor, and that the letters "IR" are hardly perceptible in
the image, even at the front interface. As discussed in section
\ref{secdimL}, the reason for this phenomenon is related with the
strong reflections from the rod medium slab due to the resonant
excitation of surface waves.

At higher frequencies, for example at 30THz and 31THz, see Fig.
\ref{30} and Fig. \ref{31}, the imaging properties are excellent. In
this frequency range the surface wave mode cannot be excited and so
there is little interaction between the transmission device and the
source. Due to this reason the "IR" shaped letters are sharply
defined in the near-field distribution at the source plane (Fig.
\ref{30} and Fig. \ref{31}). Also, consistently with our theoretical
model, the rod transmission device is able to canalize the image
from the front interface to the back interface with subwavelength
resolution. The best imaging occurs at 31THz. We can estimate the
resolution (minimum distance that makes two point sources resolved)
of the transmission device as explained next. From Fig.
\ref{frqStudy} it is seen that at 30THz, $T>0.7$ (half-power
criterion) for $k_{||}<3.43 \beta$, or equivalently for $k_{||}< 5.1
\beta_0$. Thus, the resolution of the transmission device is about
5.1 times better than that of free-space, i.e. the resolution is
around $\lambda_0 / (2 \times 5.1) = \lambda_0 /10.2$. This rough
criterion is consistent with the resolution estimated directly from
the image. It can be checked that the thickness of the stem of
letter "I" at the back interface is approximately $5a \approx
\lambda_0/9.3$. This implies that the resolution is around
$\lambda_0/9.3$.

For frequencies larger than 31THz the imaging properties are
slightly deteriorated, as it could be anticipated from Fig.
\ref{frqStudy}, but subwavelength resolution is still observed. This
is illustrated in Fig. \ref{33} and Fig. \ref{36} for 33THz and
36THz, respectively. At 36THz the image is not so clear as at lower
frequencies, and thus we can treat 35 THz as the upper bound of the
frequency band where imaging with good sub-wavelength resolution is
observed (the lower bound is roughly at 30 THz). It is thus clear
that the bandwidth of the proposed waveguiding mechanism is very
good: around $15\%$, unlike in other configurations proposed in the
literature (e.g. \cite{Pendry_PerfectLens}). Indeed, our conceptual
approach is very different from other works. In our structure we are
not directly exploring a resonance of the artificial material, but
instead we are only exploiting the fact that metallic rods support a
quasi-TEM dispersionless mode, which is by itself a very broadband
phenomenon.

For frequencies below 30THz the resolution and bandwidth of a
properly scaled transmission device can always be improved
\cite{SWIWM, ResolutionPRE}. This is a direct consequence of silver
becoming a better electrical conductor. It is also worth mentioning
that it is possible to get even better results at 30THz and higher
frequencies by using aluminium (Al) rods (aluminium has a larger
plasma frequency than silver), or alternatively by using rods made
of a material with an absorption peak at the frequency band of
interest. However, it is questionable if moving to higher
frequencies is meaningful because there are practical limits in the
realization of the nanorods.

\section{Conclusion \label{secconcl}}

In this work, we assessed the possibility of near-field
sub-wavelength imaging at infrared frequencies using array of thin
silver rods. Using the homogenization model proposed in
\cite{Silv_Nonlocalrods}, we studied the dimensioning of the rods
and of the array. It was discussed that the plasmonic rod medium
should be operated in the regime where it supports a quasi-TEM mode.
It was seen that subwavelength imaging is possible only if the
isofrequency contour of the quasi-TEM mode is nearly flat, and,
additionally, if the second TM-mode is strongly attenuated. In order
to fulfill these conditions and guarantee good matching with
free-space, it is necessary that the radius of the rods is larger
than the skin-depth of the metal, that the lattice constant is much
smaller than the wavelength, and that $R/a$ is small. In general it
may not be possible to attain all the conditions simultaneously, and
so a compromise may be necessary. It was demonstrated that an
additional boundary condition is necessary to calculate the
transmission characteristic of the rod medium slab using
homogenization methods. It was proved that the proposed analytical
formulation accurately characterizes the interaction of waves with
the plasmonic rods, and allows to qualitatively understand the
physics of the pertinent phenomena. To study the potentials and
possibilities of the proposed subwavelength imaging mechanism, we
numerically simulated the performance of an array of plasmonic rods
excited by an "IR" shaped near field source. Consistently with our
theoretical study, subwavelength imaging with resolution
$\lambda_0/10$ and 15\% bandwidth was demonstrated.

\bibliography{imaging}

\begin{thebibliography}{29}
\expandafter\ifx\csname natexlab\endcsname\relax\def\natexlab#1{#1}\fi
\expandafter\ifx\csname bibnamefont\endcsname\relax
  \def\bibnamefont#1{#1}\fi
\expandafter\ifx\csname bibfnamefont\endcsname\relax
  \def\bibfnamefont#1{#1}\fi
\expandafter\ifx\csname citenamefont\endcsname\relax
  \def\citenamefont#1{#1}\fi
\expandafter\ifx\csname url\endcsname\relax
  \def\url#1{\texttt{#1}}\fi
\expandafter\ifx\csname urlprefix\endcsname\relax\def\urlprefix{URL }\fi
\providecommand{\bibinfo}[2]{#2}
\providecommand{\eprint}[2][]{\url{#2}}

\bibitem[{\citenamefont{Pendry}(2000)}]{Pendry_PerfectLens}
\bibinfo{author}{\bibfnamefont{J.}~\bibnamefont{Pendry}},
  \bibinfo{journal}{Phys. Rev. Lett.} \textbf{\bibinfo{volume}{85}},
  \bibinfo{pages}{3966} (\bibinfo{year}{2000}).

\bibitem[{\citenamefont{Garcia and
  Nieto-Vesperinas}(2002)}]{Garcia_LimitPendry}
\bibinfo{author}{\bibfnamefont{N.}~\bibnamefont{Garcia}} \bibnamefont{and}
  \bibinfo{author}{\bibfnamefont{M.}~\bibnamefont{Nieto-Vesperinas}},
  \bibinfo{journal}{Phys. Rev. Lett.} \textbf{\bibinfo{volume}{88}},
  \bibinfo{pages}{207403} (\bibinfo{year}{2002}).

\bibitem[{\citenamefont{Smith et~al.}(2003)\citenamefont{Smith, Schurig,
  Rosebluth, Schultz, Anantha-Ramakrishna, and Pendry}}]{Smith_LimitPendry}
\bibinfo{author}{\bibfnamefont{D.~R.} \bibnamefont{Smith}},
  \bibinfo{author}{\bibfnamefont{D.}~\bibnamefont{Schurig}},
  \bibinfo{author}{\bibfnamefont{M.}~\bibnamefont{Rosebluth}},
  \bibinfo{author}{\bibfnamefont{S.}~\bibnamefont{Schultz}},
  \bibinfo{author}{\bibfnamefont{S.}~\bibnamefont{Anantha-Ramakrishna}},
  \bibnamefont{and} \bibinfo{author}{\bibfnamefont{J.~B.}
  \bibnamefont{Pendry}}, \bibinfo{journal}{Appl. Phys. Lett.}
  \textbf{\bibinfo{volume}{82}}, \bibinfo{pages}{1506} (\bibinfo{year}{2003}).

\bibitem[{\citenamefont{Podolskiy and Narimanov}(2005)}]{Podolskiy}
\bibinfo{author}{\bibfnamefont{V.~A.} \bibnamefont{Podolskiy}}
  \bibnamefont{and} \bibinfo{author}{\bibfnamefont{E.~E.}
  \bibnamefont{Narimanov}}, \bibinfo{journal}{Opt. Lett.}
  \textbf{\bibinfo{volume}{30}}, \bibinfo{pages}{75} (\bibinfo{year}{2005}).

\bibitem[{\citenamefont{Luo et~al.}(2002)\citenamefont{Luo, Johnson,
  Joannopoulos, and Pendry}}]{Luo_PC}
\bibinfo{author}{\bibfnamefont{C.}~\bibnamefont{Luo}},
  \bibinfo{author}{\bibfnamefont{S.~G.} \bibnamefont{Johnson}},
  \bibinfo{author}{\bibfnamefont{J.~D.} \bibnamefont{Joannopoulos}},
  \bibnamefont{and} \bibinfo{author}{\bibfnamefont{J.}~\bibnamefont{Pendry}},
  \bibinfo{journal}{Phys. Rev. B} \textbf{\bibinfo{volume}{65}},
  \bibinfo{pages}{201104} (\bibinfo{year}{2002}).

\bibitem[{\citenamefont{Luo et~al.}(2003)\citenamefont{Luo, Johnson,
  Joannopoulos, and Pendry}}]{Luo_PC_2}
\bibinfo{author}{\bibfnamefont{C.}~\bibnamefont{Luo}},
  \bibinfo{author}{\bibfnamefont{S.~G.} \bibnamefont{Johnson}},
  \bibinfo{author}{\bibfnamefont{J.~D.} \bibnamefont{Joannopoulos}},
  \bibnamefont{and} \bibinfo{author}{\bibfnamefont{J.}~\bibnamefont{Pendry}},
  \bibinfo{journal}{Phys. Rev. B} \textbf{\bibinfo{volume}{68}},
  \bibinfo{pages}{045115} (\bibinfo{year}{2003}).

\bibitem[{\citenamefont{Freire and Marques}(2005)}]{Marques_magnetoLens}
\bibinfo{author}{\bibfnamefont{M.~J.} \bibnamefont{Freire}} \bibnamefont{and}
  \bibinfo{author}{\bibfnamefont{R.}~\bibnamefont{Marques}},
  \bibinfo{journal}{Appl. Phys. Lett.} \textbf{\bibinfo{volume}{86}},
  \bibinfo{pages}{182505} (\bibinfo{year}{2005}).

\bibitem[{\citenamefont{Li and Lin}(2003)}]{Li_PC}
\bibinfo{author}{\bibfnamefont{Z.-Y.} \bibnamefont{Li}} \bibnamefont{and}
  \bibinfo{author}{\bibfnamefont{L.-L.} \bibnamefont{Lin}},
  \bibinfo{journal}{Phys. Rev. B} \textbf{\bibinfo{volume}{68}},
  \bibinfo{pages}{245110} (\bibinfo{year}{2003}).

\bibitem[{\citenamefont{Chien et~al.}(2004)\citenamefont{Chien, Tang, Kuo,
  Chen, and Ye}}]{Chien_PC}
\bibinfo{author}{\bibfnamefont{H.-T.} \bibnamefont{Chien}},
  \bibinfo{author}{\bibfnamefont{H.-T.} \bibnamefont{Tang}},
  \bibinfo{author}{\bibfnamefont{C.-H.} \bibnamefont{Kuo}},
  \bibinfo{author}{\bibfnamefont{C.-C.} \bibnamefont{Chen}}, \bibnamefont{and}
  \bibinfo{author}{\bibfnamefont{Z.}~\bibnamefont{Ye}}, \bibinfo{journal}{Phys.
  Rev. B} \textbf{\bibinfo{volume}{70}}, \bibinfo{pages}{113101}
  (\bibinfo{year}{2004}).

\bibitem[{\citenamefont{Kuo and Ye}(2004)}]{Kuo_PC}
\bibinfo{author}{\bibfnamefont{C.-H.} \bibnamefont{Kuo}} \bibnamefont{and}
  \bibinfo{author}{\bibfnamefont{Z.}~\bibnamefont{Ye}}, \bibinfo{journal}{Phys.
  Rev. E} \textbf{\bibinfo{volume}{70}}, \bibinfo{pages}{056608}
  (\bibinfo{year}{2004}).

\bibitem[{\citenamefont{Zhang}(2004)}]{Zhang_PC}
\bibinfo{author}{\bibfnamefont{X.}~\bibnamefont{Zhang}},
  \bibinfo{journal}{Phys. Rev. B} \textbf{\bibinfo{volume}{70}},
  \bibinfo{pages}{205102} (\bibinfo{year}{2004}).

\bibitem[{\citenamefont{Belov et~al.}(2005)\citenamefont{Belov, Simovski, and
  Ikonen}}]{canal}
\bibinfo{author}{\bibfnamefont{P.~A.} \bibnamefont{Belov}},
  \bibinfo{author}{\bibfnamefont{C.~R.} \bibnamefont{Simovski}},
  \bibnamefont{and} \bibinfo{author}{\bibfnamefont{P.}~\bibnamefont{Ikonen}},
  \bibinfo{journal}{Phys.Rev. B.} \textbf{\bibinfo{volume}{71}},
  \bibinfo{pages}{193105} (\bibinfo{year}{2005}).

\bibitem[{\citenamefont{Ikonen et~al.}(2006)\citenamefont{Ikonen, Belov,
  Simovski, and Maslovski}}]{canal_CWM}
\bibinfo{author}{\bibfnamefont{P.}~\bibnamefont{Ikonen}},
  \bibinfo{author}{\bibfnamefont{P.}~\bibnamefont{Belov}},
  \bibinfo{author}{\bibfnamefont{C.}~\bibnamefont{Simovski}}, \bibnamefont{and}
  \bibinfo{author}{\bibfnamefont{S.}~\bibnamefont{Maslovski}},
  \bibinfo{journal}{Phys. Rev. B.} \textbf{\bibinfo{volume}{73}},
  \bibinfo{pages}{073102} (\bibinfo{year}{2006}).

\bibitem[{\citenamefont{Belov et~al.}(2006)\citenamefont{Belov, Hao, and
  Sudhakaran}}]{SWIWM}
\bibinfo{author}{\bibfnamefont{P.~A.} \bibnamefont{Belov}},
  \bibinfo{author}{\bibfnamefont{Y.}~\bibnamefont{Hao}}, \bibnamefont{and}
  \bibinfo{author}{\bibfnamefont{S.}~\bibnamefont{Sudhakaran}},
  \bibinfo{journal}{Phys. Rev. B.} \textbf{\bibinfo{volume}{73}},
  \bibinfo{pages}{033108} (\bibinfo{year}{2006}).

\bibitem[{\citenamefont{Belov and Silveirinha}(2006)}]{ResolutionPRE}
\bibinfo{author}{\bibfnamefont{P.~A.} \bibnamefont{Belov}} \bibnamefont{and}
  \bibinfo{author}{\bibfnamefont{M.}~\bibnamefont{Silveirinha}},
  \bibinfo{journal}{Phys. Rev. E} \textbf{\bibinfo{volume}{73}},
  \bibinfo{pages}{056607} (\bibinfo{year}{2006}).

\bibitem[{\citenamefont{Ono et~al.}(2005)\citenamefont{Ono, ichi Kato, and
  Kawata}}]{Ono}
\bibinfo{author}{\bibfnamefont{A.}~\bibnamefont{Ono}},
  \bibinfo{author}{\bibfnamefont{J.}~\bibnamefont{ichi Kato}},
  \bibnamefont{and} \bibinfo{author}{\bibfnamefont{S.}~\bibnamefont{Kawata}},
  \bibinfo{journal}{Phys. Rev. Lett.} \textbf{\bibinfo{volume}{95}},
  \bibinfo{pages}{267407} (\bibinfo{year}{2005}).

\bibitem[{\citenamefont{Silveirinha}(2006{\natexlab{a}})}]{Silv_Nonlocalrods}
\bibinfo{author}{\bibfnamefont{M.}~\bibnamefont{Silveirinha}},
  \bibinfo{journal}{Phys. Rev. E} \textbf{\bibinfo{volume}{73}},
  \bibinfo{pages}{046612} (\bibinfo{year}{2006}{\natexlab{a}}).

\bibitem[{\citenamefont{Mittleman}(2002)}]{SensingTHz}
\bibinfo{author}{\bibfnamefont{D.~E.} \bibnamefont{Mittleman}},
  \emph{\bibinfo{title}{Sensing with Terahertz Radiation}}
  (\bibinfo{publisher}{Springer, Heidelberg}, \bibinfo{year}{2002}).

\bibitem[{\citenamefont{Belov et~al.}(2003)\citenamefont{Belov, Marques,
  Maslovski, Nefedov, Silveirinha, Simovski, and Tretyakov}}]{WMPRB}
\bibinfo{author}{\bibfnamefont{P.}~\bibnamefont{Belov}},
  \bibinfo{author}{\bibfnamefont{R.}~\bibnamefont{Marques}},
  \bibinfo{author}{\bibfnamefont{S.}~\bibnamefont{Maslovski}},
  \bibinfo{author}{\bibfnamefont{I.}~\bibnamefont{Nefedov}},
  \bibinfo{author}{\bibfnamefont{M.}~\bibnamefont{Silveirinha}},
  \bibinfo{author}{\bibfnamefont{C.}~\bibnamefont{Simovski}}, \bibnamefont{and}
  \bibinfo{author}{\bibfnamefont{S.}~\bibnamefont{Tretyakov}},
  \bibinfo{journal}{Phys. Rev. B} \textbf{\bibinfo{volume}{67}},
  \bibinfo{pages}{113103} (\bibinfo{year}{2003}).

\bibitem[{\citenamefont{Ordal et~al.}(1985)\citenamefont{Ordal, Bell, Jr, Long,
  and Querry}}]{OptConst1}
\bibinfo{author}{\bibfnamefont{M.~A.} \bibnamefont{Ordal}},
  \bibinfo{author}{\bibfnamefont{R.~J.} \bibnamefont{Bell}},
  \bibinfo{author}{\bibfnamefont{R.~W.~A.} \bibnamefont{Jr}},
  \bibinfo{author}{\bibfnamefont{L.~L.} \bibnamefont{Long}}, \bibnamefont{and}
  \bibinfo{author}{\bibfnamefont{M.~R.} \bibnamefont{Querry}},
  \bibinfo{journal}{Applied Optics} \textbf{\bibinfo{volume}{24}},
  \bibinfo{pages}{4493} (\bibinfo{year}{1985}).

\bibitem[{\citenamefont{Ordal et~al.}(1983)\citenamefont{Ordal, Long, Bell,
  Bell, Bell, Jr., and Ward}}]{OptConst2}
\bibinfo{author}{\bibfnamefont{M.~A.} \bibnamefont{Ordal}},
  \bibinfo{author}{\bibfnamefont{L.~L.} \bibnamefont{Long}},
  \bibinfo{author}{\bibfnamefont{R.~J.} \bibnamefont{Bell}},
  \bibinfo{author}{\bibfnamefont{S.~E.} \bibnamefont{Bell}},
  \bibinfo{author}{\bibfnamefont{R.~R.} \bibnamefont{Bell}},
  \bibinfo{author}{\bibfnamefont{R.~W.~A.} \bibnamefont{Jr.}},
  \bibnamefont{and} \bibinfo{author}{\bibfnamefont{C.~A.} \bibnamefont{Ward}},
  \bibinfo{journal}{Applied Optics} \textbf{\bibinfo{volume}{22}},
  \bibinfo{pages}{1099} (\bibinfo{year}{1983}).

\bibitem[{\citenamefont{Johnson and Christy}(1972)}]{OptConst3}
\bibinfo{author}{\bibfnamefont{P.~B.} \bibnamefont{Johnson}} \bibnamefont{and}
  \bibinfo{author}{\bibfnamefont{R.~W.} \bibnamefont{Christy}},
  \bibinfo{journal}{Phys. Rev B} \textbf{\bibinfo{volume}{6}},
  \bibinfo{pages}{4370} (\bibinfo{year}{1972}).

\bibitem[{\citenamefont{El-Kady et~al.}(2000)\citenamefont{El-Kady, Sigalas,
  Biswas, Ho, and Soukoulis}}]{MetalPC}
\bibinfo{author}{\bibfnamefont{I.}~\bibnamefont{El-Kady}},
  \bibinfo{author}{\bibfnamefont{M.~M.} \bibnamefont{Sigalas}},
  \bibinfo{author}{\bibfnamefont{R.}~\bibnamefont{Biswas}},
  \bibinfo{author}{\bibfnamefont{K.~M.} \bibnamefont{Ho}}, \bibnamefont{and}
  \bibinfo{author}{\bibfnamefont{C.~M.} \bibnamefont{Soukoulis}},
  \bibinfo{journal}{Phys. Rev B} \textbf{\bibinfo{volume}{62}},
  \bibinfo{pages}{15299} (\bibinfo{year}{2000}).

\bibitem[{\citenamefont{Wang and Mittleman}(2004)}]{WG_THZ_Nature}
\bibinfo{author}{\bibfnamefont{K.}~\bibnamefont{Wang}} \bibnamefont{and}
  \bibinfo{author}{\bibfnamefont{D.~M.} \bibnamefont{Mittleman}},
  \bibinfo{journal}{Nature (London)} \textbf{\bibinfo{volume}{432}},
  \bibinfo{pages}{376} (\bibinfo{year}{2004}).

\bibitem[{\citenamefont{Cao and Jahns}(2005)}]{WG_THZ_1}
\bibinfo{author}{\bibfnamefont{Q.}~\bibnamefont{Cao}} \bibnamefont{and}
  \bibinfo{author}{\bibfnamefont{J.}~\bibnamefont{Jahns}},
  \bibinfo{journal}{Optics Express} \textbf{\bibinfo{volume}{13}},
  \bibinfo{pages}{511} (\bibinfo{year}{2005}).

\bibitem[{\citenamefont{Takahara et~al.}(1997)\citenamefont{Takahara,
  Yamagishi, Taki, Morimoto, and Kobayashi}}]{ENG_WG_Tak}
\bibinfo{author}{\bibfnamefont{J.}~\bibnamefont{Takahara}},
  \bibinfo{author}{\bibfnamefont{S.}~\bibnamefont{Yamagishi}},
  \bibinfo{author}{\bibfnamefont{H.}~\bibnamefont{Taki}},
  \bibinfo{author}{\bibfnamefont{A.}~\bibnamefont{Morimoto}}, \bibnamefont{and}
  \bibinfo{author}{\bibfnamefont{T.}~\bibnamefont{Kobayashi}},
  \bibinfo{journal}{Optics Letters} \textbf{\bibinfo{volume}{22}},
  \bibinfo{pages}{475} (\bibinfo{year}{1997}).

\bibitem[{\citenamefont{Agranovich and V.Ginzburg}(1966)}]{Agranovich}
\bibinfo{author}{\bibfnamefont{V.}~\bibnamefont{Agranovich}} \bibnamefont{and}
  \bibinfo{author}{\bibnamefont{V.Ginzburg}}, \emph{\bibinfo{title}{Spatial
  dispersion in crystal optics and the theory of excitons}}
  (\bibinfo{publisher}{Wiley- Interscience, NY}, \bibinfo{year}{1966}).

\bibitem[{\citenamefont{Silveirinha}(2006{\natexlab{b}})}]{MarioABC}
\bibinfo{author}{\bibfnamefont{M.}~\bibnamefont{Silveirinha}},
  \bibinfo{journal}{IEEE Trans. Antennas. Propagat.}
  \textbf{\bibinfo{volume}{54}}, \bibinfo{pages}{1766}
  (\bibinfo{year}{2006}{\natexlab{b}}).

\bibitem[{CST(CST Microwave StudioTM 5.0, CST of America, Inc.,
  www.cst.com)}]{CST}
 (\bibinfo{year}{CST Microwave StudioTM 5.0, CST of America, Inc.,
  www.cst.com}).

\end{thebibliography}
\end{document}